\documentclass[twocolumn]{book}
\usepackage[dvips]{graphicx,color}
\usepackage{makeidx,universe}
\usepackage{amsmath,amssymb,bm}

\makeauthorindex

\BookTitle{Proceedings of the XXIX PHYSICS IN COLLISION}

\CopyRight{\copyright 2009 by Universal Academy Press, Inc.}

\begin{document}

\pagenumbering{arabic}

\chapter{%
{\LARGE \sf Diffractive Higgs boson photoproduction in proton-proton collisions } \\
{\normalsize \bf
M.B. Gay Ducati,
and 
G.G. Silveira} \\
{\small \it \vspace{-.5\baselineskip}
High Energy Physics Phenomenology Group - GFPAE, Instituto de F\'{i}sica, Universidade Federal do Rio Grande do Sul \\
Av. Bento Gon\c{c}alves, 9500 - Caixa Postal 15051, CEP 91501-970 - Porto Alegre, RS, Brazil.
}
}

\AuthorContents{G.G.\ Silveira, and M.B.\ Gay Ducati} 

\AuthorIndex{Silveira}{G.G.} 
\AuthorIndex{Gay Ducati}{M.B.}

  \baselineskip=10pt
  \parindent=10pt

\section*{Abstract}

We propose a new approach for the Higgs boson production by Double Pomeron Exchange (DPE) in the Deeply Virtual Compton Scattering, where a color dipole interacts diffractively with the proton by DPE. Applying it to Peripheral Collisions, we predict a cross section around 0.1 fb, which is similar to that obtained from the $\gamma\gamma$ subprocess. Although this result is lower than the prediction from the KMR approach, our results are competitive with the $\gamma\gamma$ and $I\hspace{-3pt}PI\hspace{-3pt}P$ subprocesses with a more precise proposal for the Gap Survival Probability.

\section{Introduction}

The Higgs boson production by Double Pomeron Exchange (DPE) was proposed to detect it in diffractive processes \cite{KMR1997}. In Peripheral Collisions, we propose the application of the DPE in the $\gamma p$ subprocess to centrally produce the Higgs boson \cite{peiHiggs}. In this case, the interaction will occur between the colliding proton and the emitted photon from the electromagnetic field around the second proton \cite{hencken}. This process is described by a DVCS-like approach, where the gluons are exchanged in the $t$-channel of the photon-proton subprocess. This approach allows the introduction of the impact factor formalism to describe the splitting of the photon in a color dipole \cite{peiHiggs}, and to predict the production cross section.

\section{Partonic Level}

\begin{figure}[t]
   \begin{center}
      \includegraphics[width=.38\textwidth]{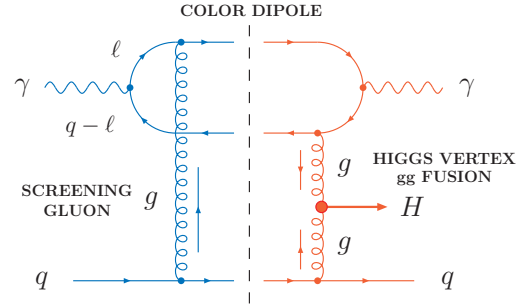}
       \caption{Feynman diagram representing the Higgs boson photoproduction by DPE.}
    \label{fig1}
   \end{center}
\end{figure}

Considering the partonic process $\gamma q \to \gamma + H + q$, one can compute the amplitude of the splitting photon, and its interaction with the quark inside the proton. The Fig.(\ref{fig1}) shows the diagram of the $\gamma q$ process, which has four different contributions coming from the coupling of the gluons to the dipole. Moreover, the amplitude is calculated using the Cutkosky rules, where a central line cuts the diagram in two parts. The imaginary part of the amplitude can be computed by
\begin{eqnarray}
{\rm{Im}}A = \frac{1}{2} \int d(PS)_{3} \, {\cal{A}}_{L} \, {\cal{A}}_{R},
\label{cut-rule}
\end{eqnarray}
where ${\cal{A}}_{i}$ is the amplitude of each side of the cut, and $d(PS)_{3}$ is the volume element of the three-body phase space. Using the Feynman rules, the scattering amplitude for a transversal polarized photon is given by (for more details, see Ref.\cite{peiHiggs})
\begin{eqnarray}\nonumber
(\textrm{Im}A)_{T} = - \frac{s}{3} \left( \frac{M^{2}_{H}}{\pi v}  \right) \alpha_{s}^{3} \, \alpha_{_{\textrm{em}}} \sum_{q} e^{2}_{q} \left( \frac{2C_{F}}{N_{c}} \right) \int \frac{d\boldsymbol{k}^{2}}{\boldsymbol{k}^{6}} \\
\times \left[ \int_{0}^{1} \frac{\boldsymbol{k}^{2}[\tau^{2} + (1 - \tau)^{2}][\alpha^{2} + (1 - \alpha)^{2}]}{\boldsymbol{k}^{2} \tau (1 - \tau) + Q^{2} \alpha(1 - \alpha)} \; d\alpha d\tau \right]\!,
\label{amp-im-1}
\end{eqnarray}
where $N_{c}$ is the color number, $v = 246 \; \textrm{GeV}$ is the v.e.v. of the Electroweak Theory, $\alpha_{s}$ and $\alpha_{em}$ are the coupling constant of QCD and QED, respectively, $C_{F}=(N^{2}_{c}-1)/2N_{c}$, and $\sum e^{2}_{q}$ is the sum of the electric charges of the quark $u$, $d$, and $s$. Finally, the event rate for central rapidity ($y_{_{H}}=0$), and no exchanged momentum by the dipole ({{$t= -{q}^{2}_{\perp} =0$}}), reads
\begin{eqnarray}\nonumber
\!\!\!\!\!\!\frac{d\sigma}{dy_{_{H}}d\boldsymbol{p}^{2}dt} \!\!\! &=& \!\!\! \frac{1}{162 \pi^{4}} \left( \frac{M^{2}_{H}}{N_{c}v} \right)^{2} \alpha_{s}^{4} \, \alpha_{_{\textrm{em}}}^{2} \left( \sum_{q} e^{2}_{q} \right)^{2} \\
&\times& \left[ \frac{\alpha_{s} \, C_{F}}{\pi} \int \frac{d\boldsymbol{k}^{2}}{\boldsymbol{k}^{6}} \, {\cal{X}}(\boldsymbol{k}^{2},Q^{2}) \right]^{2}\!\!,
\label{eq-sec}
\end{eqnarray}
where ${\cal{X}}(\boldsymbol{k}^{2},Q^{2})$ is the function in the brackets of Eq.(\ref{amp-im-1}). Comparing this result with that obtained in the KMR approach, one sees that a sixth-order $\boldsymbol{k}$-dependence arises in the event rate due to the introduction of the color dipole in the hard subprocess.

\section{Photon-Proton Process}

To include the proton to this process, the $gq$ vertex in the $\gamma q$ subprocess should be replaced by the coupling of the gluons to the colliding proton. Then, the following replacement should be performed \cite{KMR1997}
\begin{eqnarray}
\frac{\alpha_{s} \, C_{F}}{\pi} \;\; \longrightarrow \;\; f_{g}(x,\boldsymbol{k}^{2}) = {\cal{K}} \left( \frac{\partial [xg(x,\boldsymbol{k}^{2})]}{\partial \ell n \, \boldsymbol{k}^{2}} \right),
\end{eqnarray}
where $f_{g}(x,\boldsymbol{k}^{2})$ is the non-diagonal gluon distribution function into the proton. The non-diagonality of the distribution can be approximated by a multiplicative factor ${\cal{K}}$ \cite{shuvaevetal}. However, this approximation is valid for a transfer momentum $t=0$, which can be safely adopted in this calculation if one takes a small momentum fraction, like {{$x \sim 0.01$}} \cite{KMR1997}. Then, one can  identify the gluon distribution as an unintegrated distribution function $f_{g}(x,\boldsymbol{k}^{2})$. Another phenomenological effect should be introduced to suppress the gluon radiation from the production vertex, since the screening gluon could not screen the color charge of the fusing gluons as $k \to 0$ \cite{forshaw}. In this way, we include a Sudakov form factor $e^{-S}$ in Eq.(\ref{eq-sec}) to account for the probability of non-emission in the $s$-channel.

\begin{figure}[t]
   \begin{center}
     \rotatebox{-90}{\includegraphics[width=.3\textwidth]{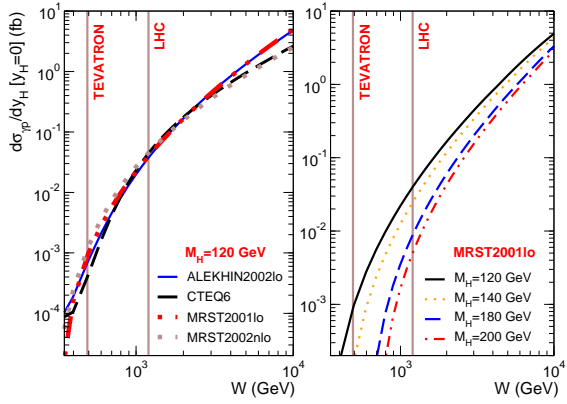}}
      \caption{Predictions for the event rate considering different parametrizations for the gluon distribution, and distinct Higgs masses. In both figures, the energies for the Tevatron and LHC are shown.}
    \label{fig1res}
   \end{center}
\end{figure}

Therefore, introducing these phenomenological aspects in Eq.(\ref{eq-sec}), and integrating it over the proton transverse momentum, the event rate for central rapidity reads
\begin{eqnarray}\nonumber
\frac{d\sigma}{dy_{_{H}}} = \frac{S^{2}_{gap}}{18 \pi^{3} b } \left( \frac{M^{2}_{H}}{N_{c}v} \right)^{2} \alpha_{s}^{4} \, \alpha_{_{\textrm{em}}}^{2} \left( \sum_{q} e^{2}_{q} \right)^{2} \int^{0.03}_{0} \!\! dq^{2}_{\perp} \\
\times \left[ \int_{\boldsymbol{k}^{2}_{0}}^{\infty} \frac{d\boldsymbol{k}^{2}}{\boldsymbol{k}^{6}} \; e^{-S(M^{2}_{H},\boldsymbol{k}^{2})} \; f_{g}(x,\boldsymbol{k}^{2}) \; {\cal{X}}(\boldsymbol{k}^{2},Q^{2}) \right]^{2}\!\!,
\label{final-eq1}
\end{eqnarray}
where $b = 5.5\mbox{ GeV}^{-2}$, and the dependence on $q^{2}_{\perp}$ appears in $Q^{2} = -k^{2}/(\gamma^{2}_{L}\beta^{2}_{L}) - q^{2}_{\perp}$ of ${\cal{X}}(\boldsymbol{k}^{2},Q^{2})$ \cite{watt}. The upper limit in the integration on $q^{2}_{\perp}$ had been chosen in order to restrict our results in the virtuality range of the real photon in Peripheral Collisions: {{$Q^{2} \lesssim 0.04$}}. Moreover, the cut $\boldsymbol{k}^{2}_{0}=0.3 \; \textrm{GeV}$ is introduced to avoid infrared divergences. Besides, $S^{2}_{gap}$ is the Gap Survival Probability, and is introduced to account the real fraction of events that will be experimentally observed. For our predictions, we adopt {{3\%}} for LHC and {{5\%}} for the Tevatron. The Fig.(\ref{fig1res}) shows the event rate for different parametrizations, and its behavior over distinct Higgs masses.

\section{Peripheral Collisions}

\begin{figure}[t]
   \begin{center}
     \rotatebox{-90}{\includegraphics[width=.3\textwidth]{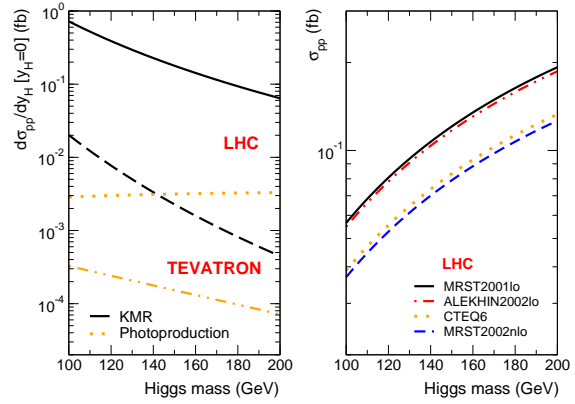}}
      \caption{The first figure shows the event rate compared to the KMR predictions, and, in the second one, the total cross section for different parametrizations for the gluon distribution.}
    \label{fig2res}
   \end{center}
\end{figure}

The hadronic cross section $\sigma_{pp}$ is calculated considering the photon emission from the colliding proton
\begin{eqnarray}
\sigma_{pp} = 2 \!\! \int_{k_{0}}^{\sqrt{s}} dk \, \frac{dn}{dk} \, \sigma_{\gamma p},
\label{pp-xsec}
\end{eqnarray}
with $dn/dk$ being the photon flux, which is given by \cite{watt}
\begin{eqnarray}\nonumber
\frac{dn}{dk} &=& \frac{\alpha_{em}}{2\pi k}\left[ 1 + \left( 1 - \frac{2k}{\sqrt{s}} \right)^{2} \right] \\
&\times& \left( \ell n A - \frac{11}{6} + \frac{3}{A} - \frac{3}{2A^{2}} + \frac{1}{3A^{2}}  \right)\!,
\label{pflux}
\end{eqnarray}
where $A \simeq 1 + (0.71 \, \textrm{GeV}^{-2})\sqrt{s}/2k^{2}$. Then, the total $\gamma p$ cross section is computed taking the integration of Eq.(\ref{final-eq1}) over the momentum fraction carried by the gluons $x=(M^{2}_{H}/W^{2})\,\textrm{exp}(\pm y_{_{H}})$. Therefore, the Higgs boson production cross section can be predicted for $pp$ collisions in the Tevatron and LHC. The Fig.(\ref{fig2res}) shows the event rate and the total cross section for its production, where, in the former, the results are compared to the ones obtained from the KMR approach \cite{forshaw}.

\section{Discussion}

The predictions show the expected behavior in comparison to the result of the KMR apporach \cite{peiHiggs}, where the event rate in the $\gamma p$ process is much larger than the direct $pp$ predictions. Furthermore, the predictions for the Tevatron also show the small possibility to produce the Higgs boson with its c.m. energy. Then, in the mass range expected to detect the Higgs boson in LHC, the KMR predictions present an event rate higher than the one from the photoproduction process, since the photon flux suppress the event rate in the high-energy regime.

\section{Conclusions}

The KMR approach suggests a GSP of 3\%, and it is applied in this work. Even so, in another approach to compute the GSP, the probability was found as 0.4\% \cite{millergap}. A comparison of these values is shown in the table below.
\begin{center}
\begin{tabular}{c|c|c}
\textbf{Subprocess} & GSP (\%) & $\sigma_{pp}$ (fb) \tabularnewline
\hline 
$I\hspace{-3pt}PI\hspace{-3pt}P$ & $\,\,$ 2.3 $\,\,$ & $\,\,$ 2.7 $\,\,$ \tabularnewline
\hline 
$I\hspace{-3pt}PI\hspace{-3pt}P$ & $\,\,$ 0.4 $\,\,$ & $\,\,$ 0.47 $\,\,$ \tabularnewline
\hline 
$\gamma\gamma$ & $\,\,$ 100 $\,\,$ & $\,\,$ 0.1 $\,\,$ \tabularnewline
\hline 
$\gamma p$ & $\,\,$ 3.0 $\,\,$ & $\,\,$ {{\textbf{0.08}}} $\,\,$ \tabularnewline
\end{tabular}
\end{center}

Although, the GSP applied in our work is similar to that of the KMR approach, it could not be appropriated to the $\gamma p$ process, since the photoproduction approach does not have the $pp$ collisions as the hard process. Therefore, we may expect a higher GSP for the Higgs boson photoproduciton, which should be studied in detail.\\ \\ This work is partially supported by CNPq.

\end{document}